\documentstyle[epsfig,aps,prc]{revtex}
\begin{document}
\draft
\title{$J/\psi$ production in relativistic heavy ion collisions from
a multi-phase transport model}
\author{Bin Zhang$^{a}$, C.M. Ko$^{b}$, Bao-An Li$^{a}$,
Zi-Wei Lin$^{b}$,
Subrata Pal$^{b}$}
\address{$^a$ Department of Chemistry and Physics,
Arkansas State University,
State University,
Arkansas 72467-0419, USA}
\address{$^b$ Cyclotron Institute and Department of Physics,
Texas A\&M University, 
College Station,
Texas 77843-3366, USA
}
\date{March 6, 2002}
\maketitle

\begin{abstract}
Using A Multi-Phase Transport (AMPT) model, we study $J/\psi$ 
production from interactions between charm and anti-charm quarks 
in initial parton phase and between $D$ and $\bar{D}$ mesons in 
final hadron phase of relativistic heavy ion collisions at the
Relativistic Heavy Ion Collider (RHIC). Including also the inverse
reactions of $J/\psi$ absorption by gluons and light mesons, we
find that the net number of $J/\psi$ from the parton and 
hadron phases is smaller than that expected from the superposition of 
initial 
nucleon-nucleon collisions, contrary to the $J/\psi$ enhancement
predicted by the kinetic formation model. The production of $J/\psi$
is further suppressed if one includes the color screening effect
in the parton phase. We have also studied the dependence of $J/\psi$
production on the charm quark mass and the effective charm meson mass.
\end{abstract}
%\newpage
%\narrowtext

\section{Introduction}

A central interest in modern nuclear physics is to produce
and study the Quark-Gluon Plasma (QGP) predicted by the
Quantum Chromodynamics (QCD) \cite{fwilczek1}. Studying 
the properties of the QGP and the deconfinement phase transition 
not only is important for understanding the QCD but also has 
astronomical implications \cite{bmuller1}.  Experiments at the
Super Proton Synchrotron (SPS) involving
collisions between heavy nuclei have shown that a large amount 
of energy is deposited around mid-rapidity, and the energy density may
already be sufficiently large for the formation of the QGP during
the initial stage of collisions \cite{uheinz1}. More recent 
experiments at RHIC at Brookhaven 
National Laboratory, where the collision energy is much higher than 
at SPS, have also shown possible effects due to the formation of
a partonic matter \cite{qm2001a}. 

To find the signal for the Quark-Gluon Plasma in relativistic
heavy ion collisions, Matsui and Satz proposed to study $J/\psi$
production in these collisions \cite{tmatsui1}. Based on the fact 
that the effective potential between charm and anti-charm quarks 
changes in the QGP due to color screening effect, no bound states can
be formed between them beyond a certain critical temperature, 
which is somewhat higher than the deconfinement temperature.
As a result, $J/\psi$ production is expected to be suppressed
if the QGP is formed in heavy ion collisions. There are other possible 
mechanisms for $J/\psi$ suppression \cite{e772a,e866a,na50a} in heavy ion 
collisions, e.g., $J/\psi$ can be destroyed by collisions with
incoming nucleons or with
gluons in the initial partonic matter \cite{eshuryak1,xxu1}
or with comoving hadrons in the hadronic matter
\cite{sgavin1,acapella1,cspieles1,cspieles2,wcassing1,jgeiss1,dkahana1,bsa1}. 
The observed abnormal suppression of $J/\psi$ in central
Pb+Pb collisions at SPS may require the formation of the 
Quark-Gluon Plasma in these collisions \cite{dkharzeev1,uheinz1}.

For heavy ion collisions at RHIC energies, unlike in previous 
fixed target experiments at SPS, multiple pairs of charm-anti-charm 
quarks can be produced in one event, and these charm and anti-charm 
quarks can interact and produce the $J/\psi$ \cite{rthews1}. 
This mechanism is the inverse reaction of $J/\psi$ absorption 
by gluons. Using a kinetic approach, it was found that despite 
$J/\psi$ dissociation due to color screening, this reaction 
is important in the quark-gluon plasma with temperatures between 
those for $J/\psi$ dissociation and the deconfinement, and
would lead to an enhanced production of $J/\psi$ in heavy ion
collisions at RHIC. A possible signal for QGP at RHIC has thus been 
changed from $J/\psi$ suppression to $J/\psi$ enhancement. 
The statistical fragmentation model 
\cite{pbraunmunzinger1,pbraunmunzinger2,lgrandchamp1}, which has
been extensively used for studying particle production in heavy 
ion collisions at SPS energies, predicts a comparable $J/\psi$ 
production. The main assumption of this model is
that $J/\psi$ formed during hadronization of the QGP
is in chemical equilibrium with charm mesons.

In the kinetic formation model, the heavy ion collision dynamics is 
treated schematically. For a more quantitative study of this new 
$J/\psi$ production mechanism, we study 
$J/\psi$ production from both the charm-anti-charm 
quark interactions and the charm-anti-charm meson interactions
using the AMPT model \cite{bzhang1,zlin1,zlin2}.
In particular, we consider central (b=0) Au+Au collisions at the 
highest RHIC energy $\sqrt{s}=200$ AGeV. With this dynamic transport 
model, we find that the net number of produced $J/\psi$ from the parton and 
hadron phases is smaller than that expected from initial 
nucleon-nucleon collisions, contrary to the $J/\psi$ enhancement
predicted by the kinetic formation model. The production of $J/\psi$
is further suppressed if one includes the color screening effect
in the parton phase.

In Section \ref{ampt}, we give a short description of the multi-phase
transport model (AMPT). Results are presented in Section \ref{results}
for $J/\psi$ production from the initial parton phase and the 
final hadron phase. In this section, we also discuss
the dependence of $J/\psi$ production in the parton phase on the 
charm quark mass and in the hadron phase on the charm meson mass.
Finally, we summarize the paper in Section \ref{summary}.

\section{Multi-Phase Transport model and $J/\psi$ production mechanisms}
\label{ampt}

The multi-phase transport model \cite{bzhang1,zlin1,zlin2} 
that we use for the present study is a hybrid model based on
three Monte-Carlo models for the three stages of relativistic
heavy ion collisions, i.e.,  the Heavy Ion Jet 
Interaction Generator (HIJING) \cite{xwang1,mgyulassy1} for
the initial conditions, Zhang's Parton Cascade (ZPC) \cite{bzhang2}
for the parton evolution, and  A Relativistic Transport (ART) model 
\cite{bli1,bli2} for the hadron evolution.  In the HIJING model, nucleons 
are distributed according to the Woods-Saxon distribution. The parton 
distribution in a nucleon is adjusted according to the 
transverse position of the nucleon and a shadowing function based on 
the Mueller-Qiu parton recombination mechanism \cite{amueller1}. 
These partons undergo hard or semi-hard collisions and generate 
an initial parton system after the passage of two colliding 
nuclei. HIJING provides the momentum space information of the 
partons produced in the hard or semi-hard collisions.  The formation 
time of a parton is generated according to the Gyulassy-Wang model of 
parton coherent production \cite{mgyulassy2}. The position of a physical
parton is then obtained from the position of its parent nucleon by
adding a displacement given by its velocity times the formation time. 
The resulting parton phase space distribution is used as the 
initial conditions for starting the ZPC parton evolution. In the 
present version of the ZPC model, only elastic gluon-gluon scattering 
are included with a cross section of 3 mb, which is consistent 
with the estimated average screening mass in the partonic matter
formed in the collisions. When the partonic matter stops interacting, 
partons are connected back with their parent nucleons, which also suffer 
soft interactions through string excitations.  The Lund fragmentation 
model \cite{bandersson1} is then called from the HIJING model
to convert these wounded nucleons into hadrons. These hadrons 
become physical after a formation time given by 
the average freeze-out proper time of partons in a string 
plus an additional 0.7 fm/c to account for the time that is required for 
producing quark-anti-quark pairs. The ensuing hadron evolution is  
modeled by the ART model that takes into account interactions 
among nucleons, anti-nucleons, mesons, and resonances. 
For a better description of experimental hadron spectra,
the original ART model has been improved by including the 
$K^*$ meson interactions, the anti-nucleon production, and 
the $\rho$ meson interactions.

We have previously used this multi-phase transport model to study  
the suppression of primarily produced $J/\psi$ in heavy ion collisions 
at RHIC \cite{bzhang3}.  These $J/\psi$'s are produced in nucleon-nucleon 
collisions during the initial passage of two colliding nuclei. 
Because of color screening, these $J/\psi$'s are not physical ones
but correlated pairs of charm-anti-charm quarks close in phase space.
They become physical $J/\psi$'s if they remain close 
when the partonic matter cools below the $J/\psi$ dissociation temperature.
Otherwise, they will combine with nearby light quarks to form 
charm mesons at hadronization. Gluons can also destroy
the phase space correlation between the charm quark and anti-quark
pairs and hence destroy the $J/\psi$. It was found that
finite space size and lifetime have significant effects on
$J/\psi$ suppression, and different mechanisms lead to different 
suppression factors. Also, the dependence of $J/\psi$ suppression on
the size of colliding nuclei was shown to be useful for differentiating
the different suppression mechanisms. Similar results have also been 
obtained in other studies \cite{rvogt1,wcassing2}.

In the present study, we include the reactions of
$J/\psi$ production from charm-anti-charm quark interactions 
in the parton phase and from charm meson interactions in the hadron 
phase. As the primary $J/\psi$ number is much smaller than the number
of other particles participating in the equilibration, primary $J/\psi$
evolution won't affect secondary $J/\psi$ production. We do not include
primary $J/\psi$'s in the following calculations.

The charm and anti-charm quarks are generated from 
the PYTHIA model \cite{tsjostrand1}, with a cross section of
350 $\mu$b for producing a pair of charm-anti-charm pair 
in a nucleon-nucleon collision. This cross section does not include 
nuclear shadowing effect of gluons, so the 
present study gives an upper bound for charm meson and $J/\psi$
production in heavy ion collisions. In the partonic stage of
heavy ion collisions, $J/\psi$'s are then produced from charm-anti-charm 
quark interactions through the reaction ${c+\bar{c}\rightarrow J/\psi+g}$. 
The $J/\psi$ can also be destroyed by gluons through the reaction
${g+J/\psi\rightarrow c+\bar{c}}$. We further include the dissociation
of $J/\psi$ due to color screening in the partonic matter. 
To treat this process, we follow the method used in Ref. \cite{bzhang3}
by introducing a critical radius in the transverse plane of a
heavy ion collision, within which the parton energy density is
larger than the critical energy density for $J/\psi$ dissociation by 
color screening and the $J/\psi$ thus cannot be formed. 
The critical radius decreases as the partonic matter expands.
The time evolution of this critical radius in Au+Au collisions at
$\sqrt{s}=200$ AGeV has been parameterized in Ref. \cite{bzhang3}.
In addition to producing the $J/\psi$, charm and anti-charm quarks 
also undergo elastic scattering with gluons. This
elastic cross section is taken to be 3 mb in the present study.

After partons stop interacting, charm and anti-charm quarks
are converted to $D$ and $\bar{D}$ mesons. This conversion
is carried out using a delta function fragmentation scheme.
In this approach, the momentum of a charm meson
is taken to be the same as the momentum of its parent
charm quark. A formation time of 1 fm/c is also introduced 
in combining the charm quark with a nearby light quark. Because 
of the large mass of charm quark, the delta function fragmentation scheme
gives a reasonable description of charm quark fragmentation in
hadron reactions.
For simplicity, we do not differentiate between $D$ mesons and $D^*$ mesons. 
Instead, we vary the $D$ meson mass to study the influence of different 
production thresholds for $J/\psi$ production from charmed hadrons. 

In the hadronic phase, $D$ and $\bar{D}$ mesons can interact to
produce a $J/\psi$ and a light meson through the reaction
$D\bar D\to J/\psi M$, where $M$ denotes a light meson.
The $J/\psi$ can also be destroyed by light mesons through the
inverse reaction $J/\psi M\to D\bar D$. We only include reactions 
that involve $\pi$ mesons or $\rho$ mesons as they are the most abundant 
ones in the nearly baryon free matter formed in heavy ion collisions at RHIC.
The cross section for light meson destruction of $J/\psi$ is taken 
to be 3 mb as in previous transport study at SPS energies
\cite{cspieles1,cspieles2,wcassing1,jgeiss1,dkahana1,bsa1}. 
This cross section is consistent with those from 
recent theoretical studies
based on effective hadronic Lagrangians \cite{khaglin1,zlin3,asibirtsev1}, 
the quark-exchange model \cite{cywong1}, and the QCD sum rules
\cite{nielsen}. The cross section for the inverse
reaction of $J/\psi$ production from $D$ mesons is calculated 
according to the detailed balance relation. As we do not differentiate 
between $D$ mesons and $D^*$ mesons, 
the spin and isospin degrees of freedom for both $D$ 
and $D^*$ mesons are used in evaluating the inverse reaction 
cross section. This is equivalent to including
all possible combination of $D$ and $D^*$ states. We will vary
the $D$ meson mass to study the effect due to changing $J/\psi$
production threshold.

In earlier studies based on 
kinetic models \cite{ko1}, $J/\psi$ production from 
the hadron phase of relativistic heavy ion collisions was 
found to be important only at LHC \cite{ko1,pbraunmunzinger3} 
but not at RHIC as the cross section 
for $D\bar D\to J/\psi M$ used in these studies was smaller than
that used in present study. 

Although multiple pairs of charm quarks are expected to be
produced in each Au+Au collision, the number of charm quarks 
is small compared with other particles. 
For example, the rapidity density of 
charm quark pairs is about 1.5 at $\sqrt{s}=200$ AGeV. The
expected number of $J/\psi$ per event is even much smaller.
To obtain sufficient statistics for $J/\psi$ production,
charm particles and the $J/\psi$ are treated in the transport
model by the perturbative method \cite{jrandrup1,xfang1}, 
i.e., neglecting their effects on heavy ion collision dynamics. 
This method has been used extensively for studying 
rare particle production in heavy ion collisions at low energies. 
Explicitly, we include many charm quark events from primary nucleon-nucleon
interactions in the evolution of partons from a single HIJING event. 
However, only charm quarks from the same event 
are allowed to interact among themselves, and the usual method
of treating binary collisions in transport models is used in treating
these collisions. For collisions between charm and uncharm particles, 
we keep their effects on the charm particles but neglect those on
the uncharm particles. In this way, the dominant computation time for
the interactions between uncharm particles remains the same
as for a single charm event. Since there are much more uncharm particles 
than charm particles, e.g., about 300 gluons verse about 3 charm and 
anti-charm quarks per unit rapidity, neglecting
the effects of charm particles on the heavy ion collision dynamics
is expected to be a very good approximation.

\section{Results}
\label{results}

\subsection{$J/\psi$ production in parton phase}

\begin{figure}[htb]
\null\vspace{1cm}
\centerline{\epsfig{file=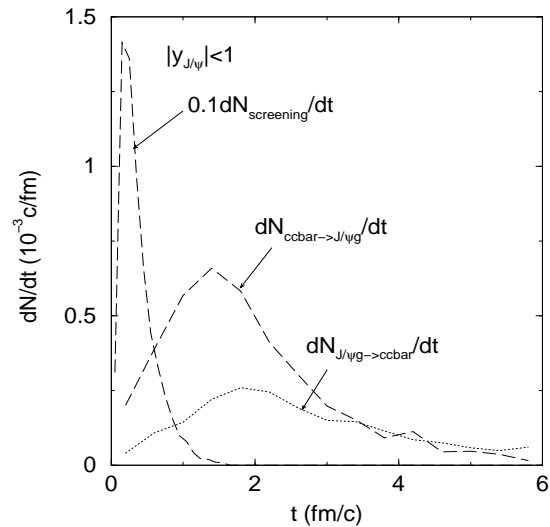,width=2.8in,height=2.8in,angle=-90}}
\null\vspace{0.5cm}
\caption{Time evolution of the production rate (dashed line) and 
the destruction rate (dotted line) per unit rapidity for
$J/\psi$ with $|y_{J/\psi}|<1$
in the parton phase with the color screening effect.
The thin dashed line denotes the collision rate per unit rapidity 
of charm and anti-charm
quarks which would have produced the $J/\psi$ if color screening effect
is absent.}
\label{fig1}
\end{figure}

We first study $J/\psi$ production in the parton phase. 
Unless otherwise indicated, the rates and numbers in the following
are all rapidity densities averaged with $|y|<1$. Fig.~\ref{fig1} gives
the time evolution of the production and destruction rates per unit rapidity
for $J/\psi$'s with $|y_{J/\psi}|<1$. 
Also shown is the rate per unit rapidity for 
collisions between charm and anti-charm quarks which would have 
produced the $J/\psi$ in the absence of color screening effect.
These unsuccessful collisions mainly occur before 0.5 fm/c when
most partonic matter have effective temperatures above the $J/\psi$
dissociation temperature. The screening effect largely diminishes after 
1 fm/c when the parton density decreases due to expansion.
At this time, the $J/\psi$ production rate starts to increase quickly,
and this is followed by an increase of the $J/\psi$ destruction rate. 
After reaching their peak values, both the production and destruction
rates decrease and become comparable at about 4 fm/c.

\begin{figure}[htb]
\null\vspace{1cm}
\centerline{\epsfig{file=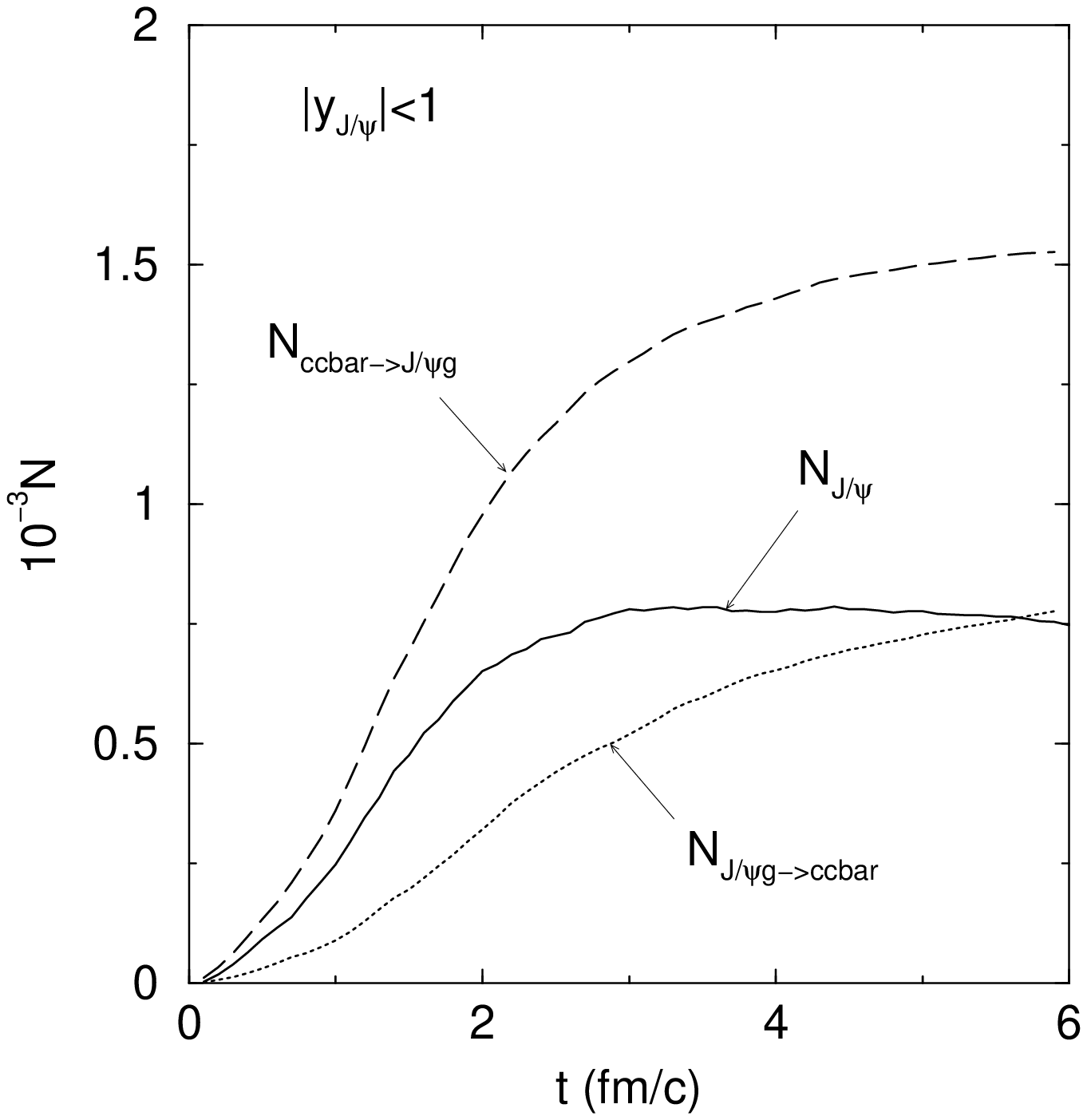,width=2.8in,height=2.8in,angle=-90}}
\null\vspace{0.5cm}
\caption{Time evolution of produced number (dashed line), destructed
number (dotted line), and net number (solid line) per unit rapidity 
for
$J/\psi$ with $|y_{J/\psi}|<1$ in the parton phase with the color
screening effect.}
\label{fig2}
\end{figure}

To see more clearly the relative importance of $J/\psi$ production
and destruction in the parton phase, we show in Fig.~\ref{fig2} the time evolution 
of the number of produced $J/\psi$ and the number of destructed $J/\psi$
as well as the net $J/\psi$ number.
The number of $J/\psi$ saturates at
about 3 fm/c. It decrease slowly afterwards as the production and
destruction rates become comparable as shown in Fig.~\ref{fig1}.

\begin{figure}[htb]
\null\vspace{1cm}
\centerline{\epsfig{file=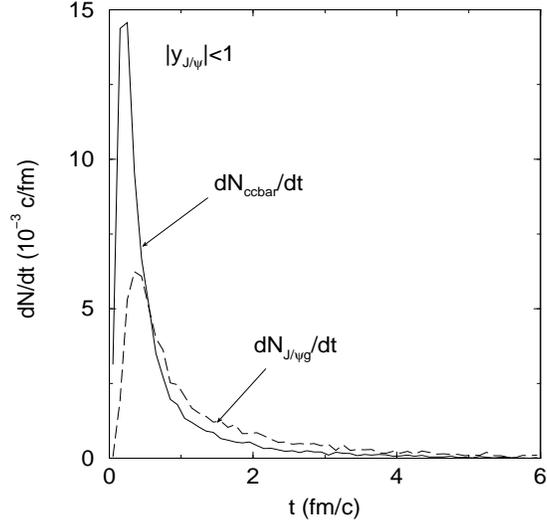,width=2.8in,height=2.8in,angle=-90}}
\null\vspace{0.5cm}
\caption{Time evolution of the production rate (solid line) and 
the destruction rate (dashed line) per unit rapidity for
$J/\psi$ with $|y_{J/\psi}|<1$
in the parton phase without color screening effect.}
\label{fig3}
\end{figure}

\begin{figure}[htb]
\null\vspace{1cm}
\centerline{\epsfig{file=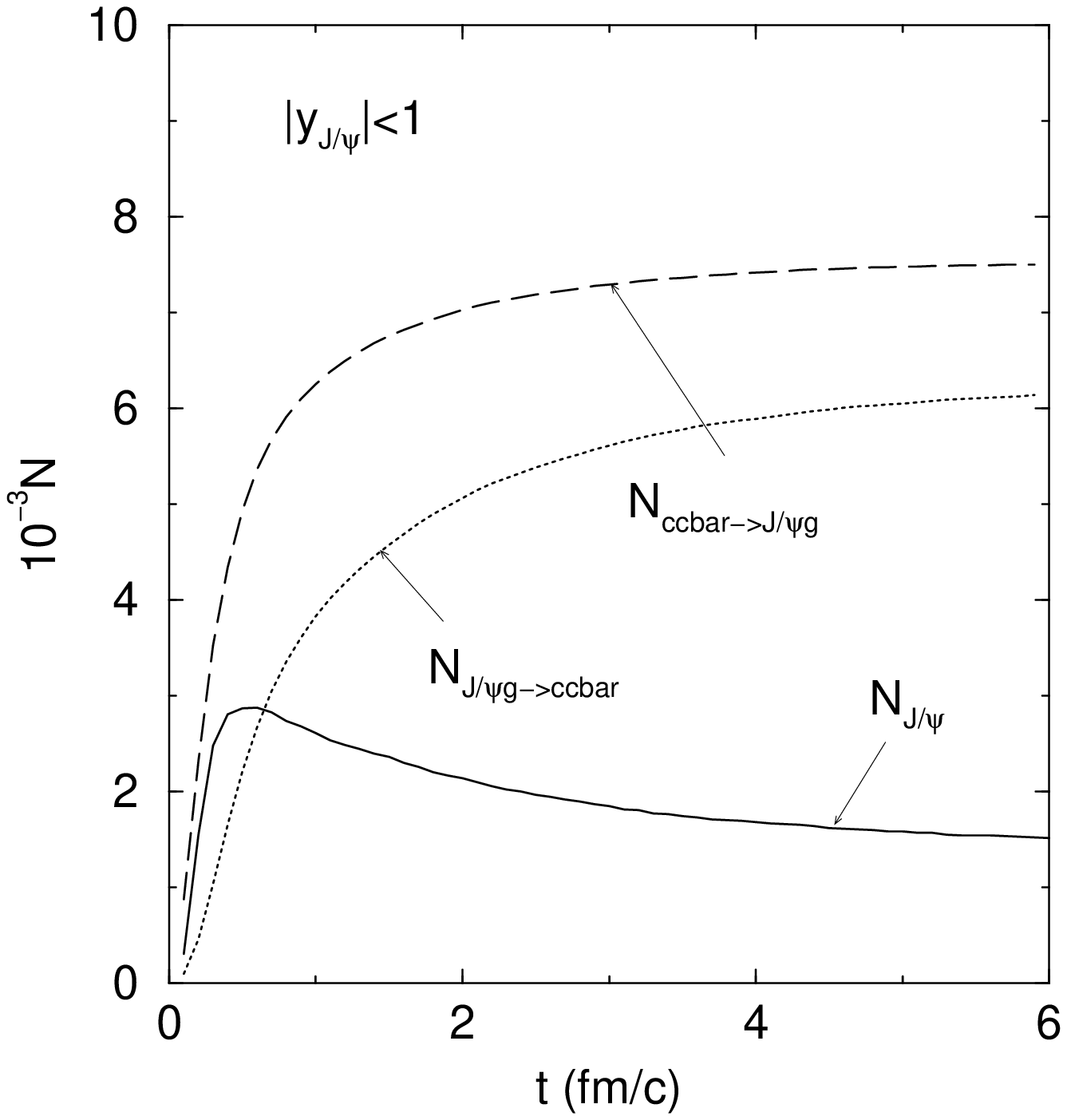,width=2.8in,height=2.8in,angle=-90}}
\null\vspace{0.5cm}
\caption{Time evolution of produced number (dashed line), destructed
number (dotted line), and net number (solid line) per unit rapidity
for
$J/\psi$ with $|y_{J/\psi}|<1$ in the parton phase without color
screening effect.}
\label{fig4}
\end{figure}

In the absence of color screening effect, collisions between charm 
and anti-charm quarks in the partonic matter that is above the
$J/\psi$ dissociation temperature can also produce the $J/\psi$.
This leads to a very high $J/\psi$ production rate before
0.5 fm/c as shown in Fig.~\ref{fig3}. The $J/\psi$ destruction rate 
during this time also reaches its peak value of about 1/3 of the maximum 
production rate. In contrast to the case including color screening
effect, the production and destruction rates become comparable
already at about 0.5 fm/c. As shown in Fig.~\ref{fig4}, these result in
a much faster increase in the numbers of produced and destructed
$J/\psi$, and an earlier peaking in the net $J/\psi$ number  
than in the case with color screening effect. Since the destruction
rate after 0.5 fm/c is still slightly larger than the production rate, 
the net $J/\psi$ number slowly decreases from 
its maximum value of about 0.003 to a final value of about 0.0014 
per event. This number is about a factor of two larger than the 
final net $J/\psi$ number for the case with color screening effect, 
which is about 0.0007 per event. 

The number of $J/\psi$ expected from primary nucleon-nucleon collisions
in Au+Au central collisions can be estimated from the $J/\psi$ production
cross section in $pp$ collisions. The cross section per unit rapidity 
at $\sqrt{s}=200$ AGeV is 0.63 $\mu$b \cite{pgavai1}. Using the nuclear 
overlap function $T_{Au+Au}(b=0)=30$ mb$^{-1}$, we expect 0.019 $J/\psi$ 
per unit rapidity to be produced in Au+Au central collisions. This is 
much larger than the predictions of 0.0007 and 0.0014 from 
the dynamical multi-phase transport model. Our study thus predicts 
a suppression of $J/\psi$ production in the parton phase of heavy ion 
collisions at RHIC, and this is opposite to the enhancement predicted
by the kinetic formation model.

\subsection{$J/\psi$ production in hadron phase}

\begin{figure}[htb]
\null\vspace{1cm}
\centerline{\epsfig{file=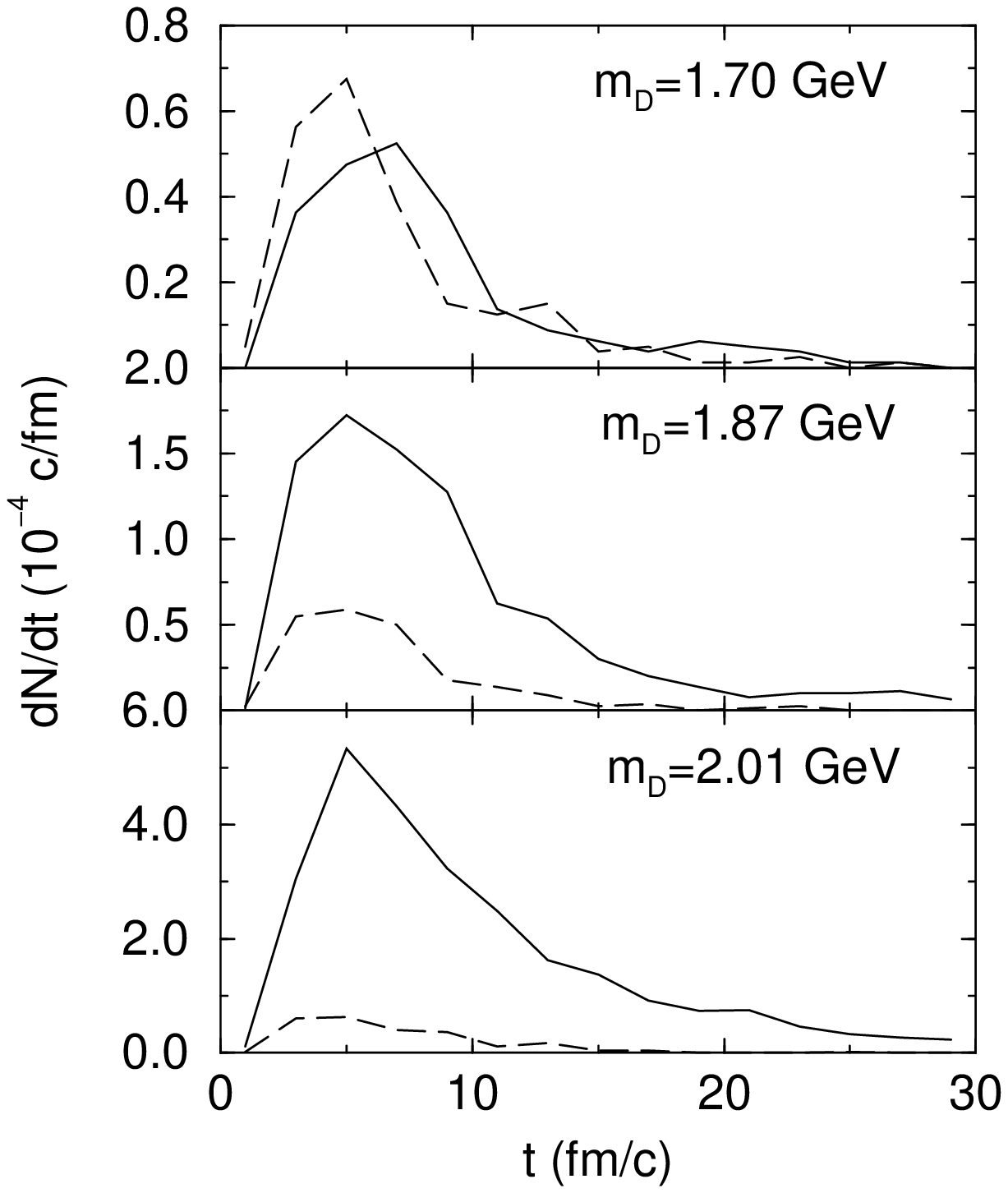,width=2.6in,height=3.3in,angle=0}}
\null\vspace{0.5cm}
\caption{Time evolution of the production rate (solid line)
and the destruction rate (dashed line) per unit rapidity 
for $J/\psi$ with $|y_{J/\psi}|<1$ 
from the hadron phase for three values of $D$ meson mass. Color screening 
effect is included in the initial parton phase.}
\label{fig5}
\end{figure}

In hadronic matter, the threshold for $J/\psi$ production is reduced 
in comparison with that in the partonic matter as the $D$ meson mass is 
larger than the charm quark mass. This threshold effect is expected 
to compensate for the lower charm density in the hadron phase than in
the parton phase.  To demonstrate this threshold effect, we 
vary the $D$ meson mass in studying $J/\psi$ production in the
hadronic matter.  Fig.~\ref{fig5} shows the time evolution of 
$J/\psi$ production and destruction rates per unit rapidity in
the hadron phase for different $D$ meson masses.  These results
are obtained with the color screening effect included in the initial 
parton phase. The upper panel is for $m_D=1.70$ GeV, which is
smaller than the $D$ meson mass in vacuum. We use this smaller
mass because theoretical studies have shown that the $D$ meson mass 
may be reduced in medium \cite{ktsushima1,ahayashigaki1,wweise1,asibirtsev2}. 
The middle panel is for $m_D=1.87$ GeV, 
which is the $D$ meson mass in vacuum. Here, we treat ${D^*}$ meson mass 
as $D$ meson mass, and the results are thus a lower bound for $J/\psi$ 
production. The lower panel is for $m_D=2.01$ GeV, which is
the ${D^*}$ mass in vacuum. Treating $D$ meson mass as ${D^*}$
meson mass in this case then gives an upper bound for $J/\psi$ production.
The destruction rates for all three $D$ meson masses are seen to have
similar magnitude. For the production rate, it is larger for
larger $D$ meson masses.

\begin{figure}[htb]
\null\vspace{1cm}
\centerline{\epsfig{file=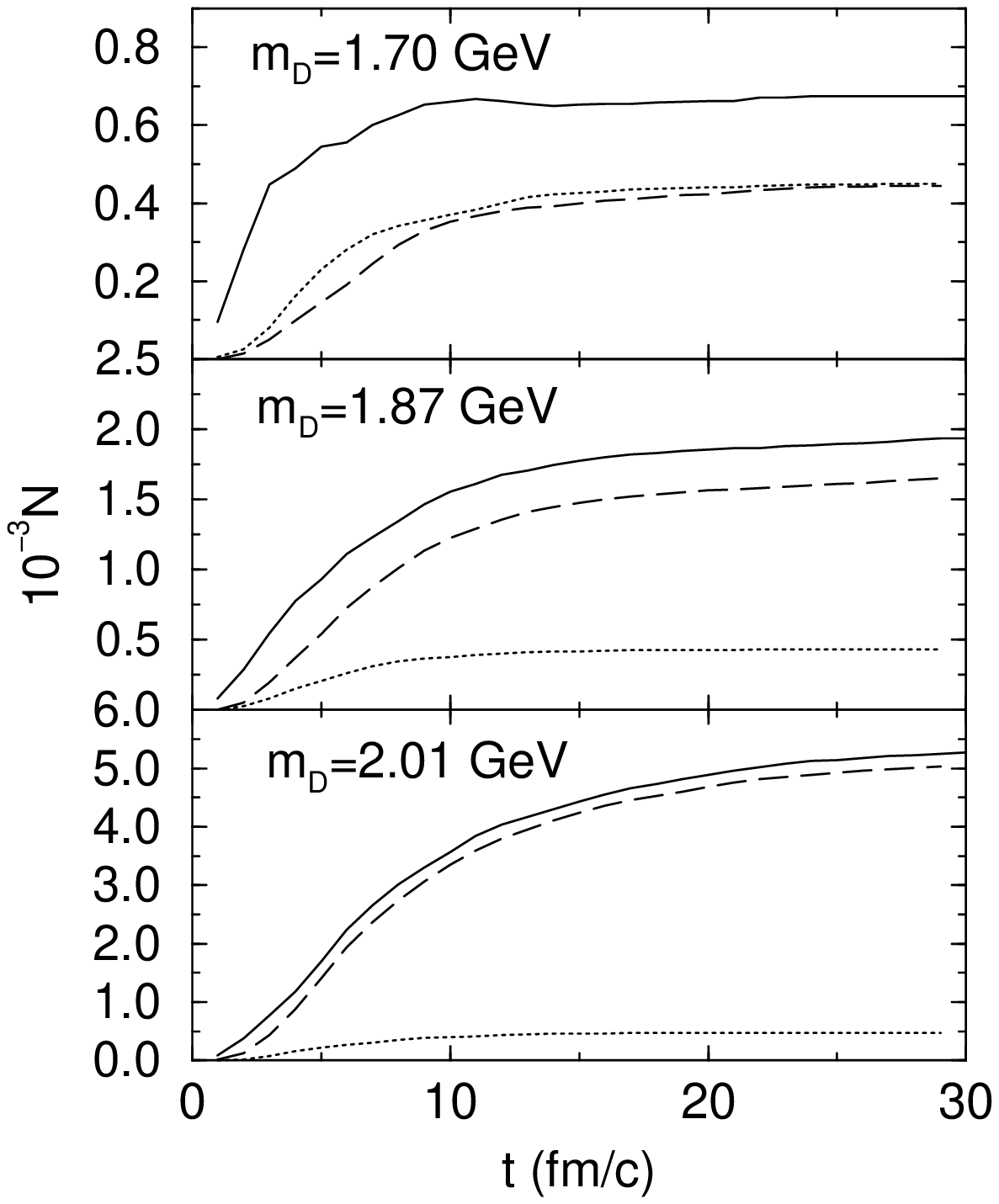,width=2.6in,height=3.3in,angle=0}}
\null\vspace{0.5cm}
\caption{Time evolution of produced number (dashed line), destructed
number (dotted line), and net number (solid line) per unit rapidity
for
$J/\psi$ with $|y_{J/\psi}|<1$ for three values of $D$ meson mass. 
Color screening effect is included in the initial parton phase.}
\label{fig6}
\end{figure}

In Fig.~\ref{fig6}, we show the time evolution of the numbers of produced and 
destructed $J/\psi$ together with that of the net $J/\psi$ number.
For $m_D=1.87$ GeV and $m_D=2.01$ GeV, the number of produced $J/\psi$ 
is always greater than the number of destructed $J/\psi$, leading
to a net production of $J/\psi$ from the hadron phase of relativistic
heavy ion collisions. With increasing $D$ meson mass, the destructed
$J/\psi$ number increases slightly while the produced $J/\psi$
increases drastically. For $m_D=1.70$ GeV, the produced and 
destructed $J/\psi$ numbers are about equal, and the final net 
$J/\psi$ number is thus about the same as that from the parton 
phase. On the other hand, for $m_D=2.01$ GeV, the final 
$J/\psi$ number equals approximately the number of produced $J/\psi$ 
from the hadron phase. The final $J/\psi$ number reflects essentially 
the one produced from the hadron phase.

Quantitatively, the final net number of $J/\psi$ per unit
rapidity averaged over $|y_{J/\psi}|<1$
is about 0.0007 per event for $m_D=1.70$ GeV and increases to 
0.0019 for $m_D=1.87$ GeV and to 0.0053 for $m_D=2.01$ GeV. 
The time evolution of the net $J/\psi$ number in the three cases 
are similar as it is largely determined by the heavy ion collision 
dynamics. The $J/\psi$ number saturates at about 15-20 fm/c, which is 
about the lifetime of the hadronic matter. 

\begin{figure}[htb]
\null\vspace{1cm}
\centerline{\epsfig{file=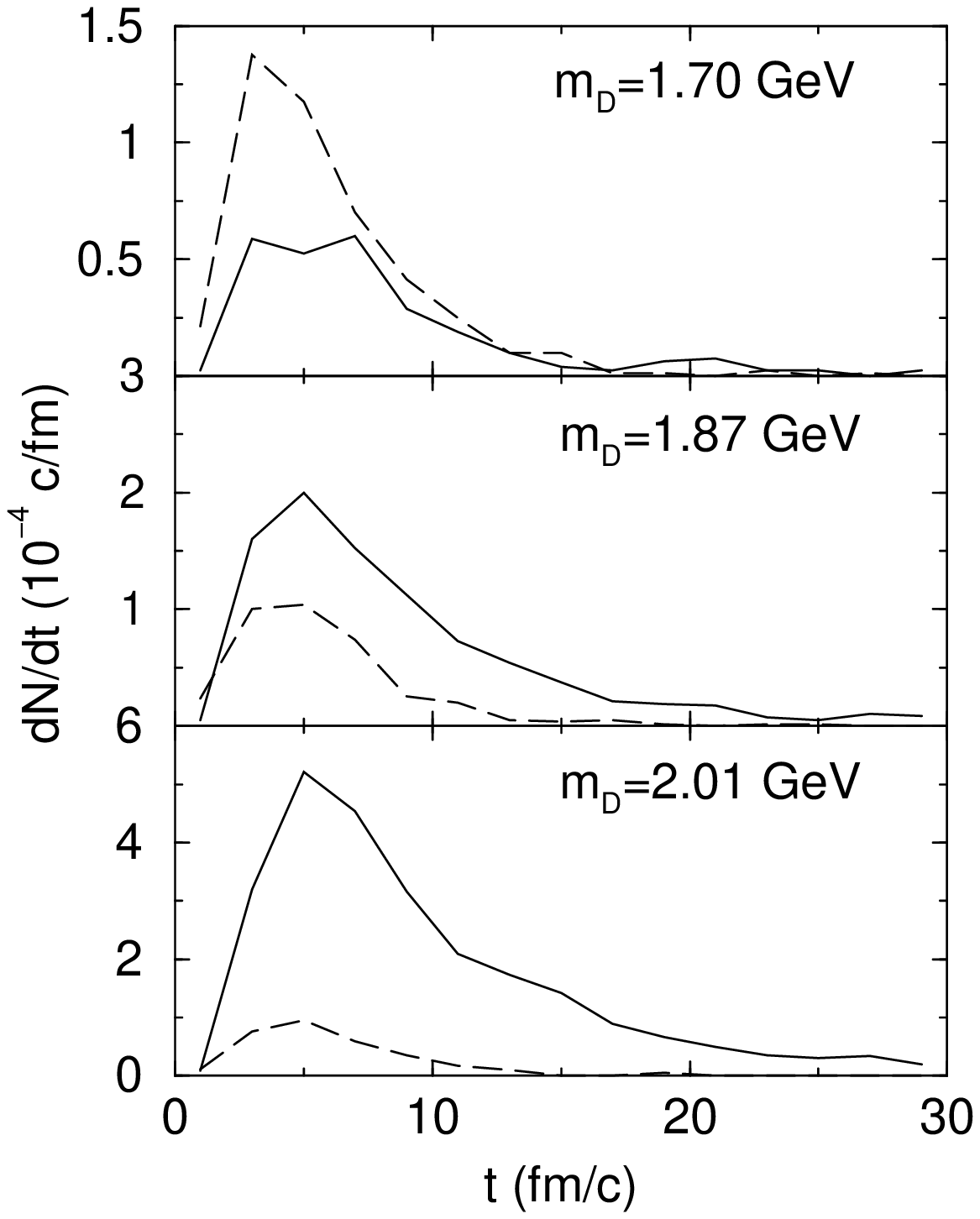,width=2.6in,height=3.3in,angle=0}}
\null\vspace{0.5cm}
\caption{Time evolution of the production rate (solid line)
and the destruction rate (dashed line) per unit rapidity 
for $J/\psi$ with $|y_{J/\psi}|<1$ 
from the hadron phase for three values of $D$ meson mass. Color screening 
effect is not included in the initial parton phase.}
\label{fig7}
\end{figure}

\begin{figure}[htb]
\null\vspace{1cm}
\centerline{\epsfig{file=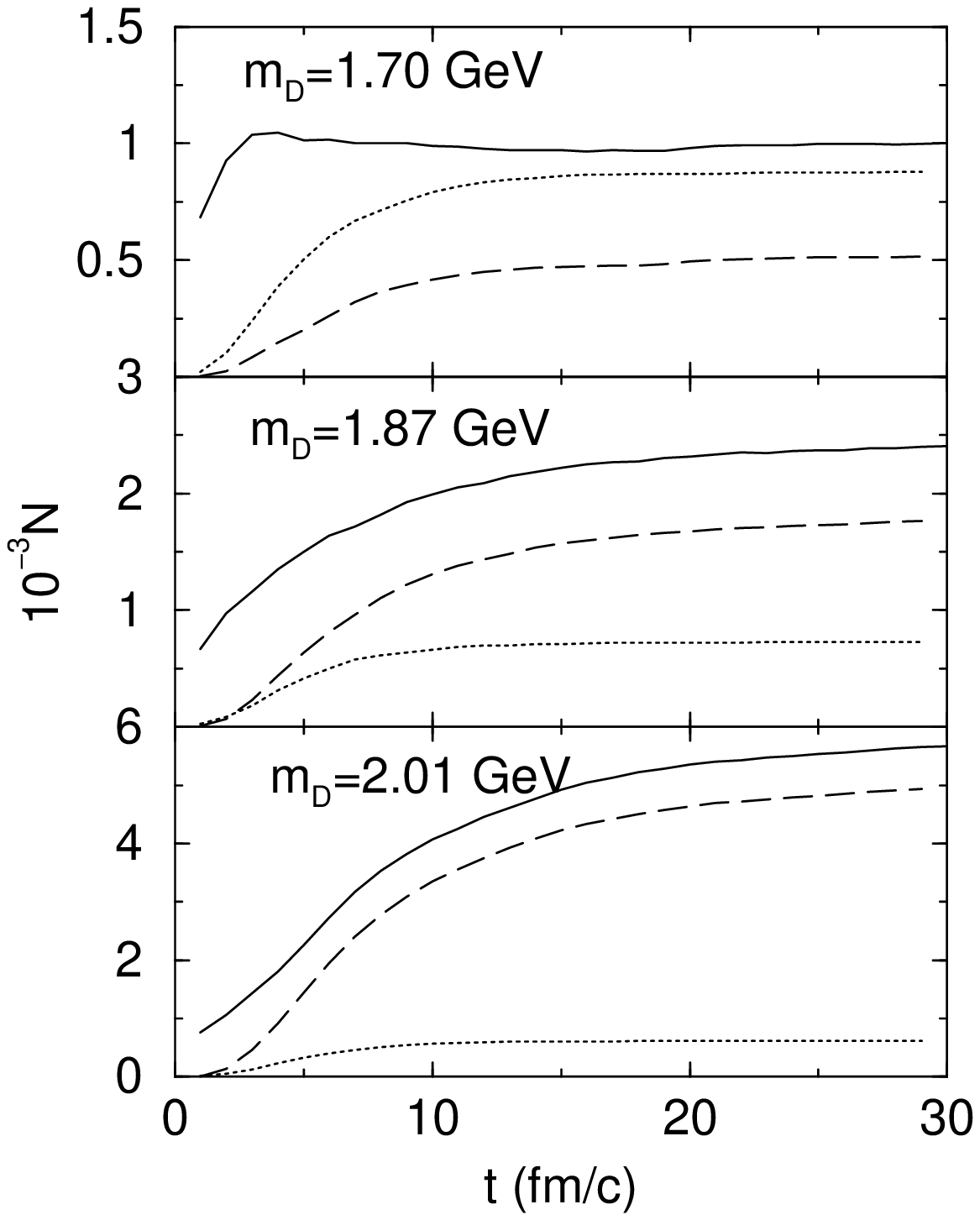,width=2.6in,height=3.3in,angle=0}}
\null\vspace{0.5cm}
\caption{Time evolution of produced number (dashed line), destructed
number (dotted line), and net number (solid line) per unit rapidity for
$J/\psi$ with $|y_{J/\psi}|<1$ in the hadron phase for three values 
of $D$ meson mass. Color screening effect is not included in the initial 
parton phase.}
\label{fig8}
\end{figure}

The results for neglecting the color screening effect in the initial
parton phase are shown in Fig.~\ref{fig7} for time evolution of the $J/\psi$
production and destruction rates and in Fig.~\ref{fig8} for the
time evolution of the numbers of produced and destructed $J/\psi$
as well as the net $J/\psi$ number.  Although there are more $J/\psi$'s
destructed in the hadronic matter, particularly for $m_D=1.70$ GeV
where the number of destructed $J/\psi$ is even larger than the 
number of produced $J/\psi$ in the hadron phase, the final
$J/\psi$ number is larger than in the case with color screening 
in the parton phase. Their numbers are 0.0010, 0.0024, and 0.0057  
for $m_D=1.70$, 1.87, and 2.01 GeV, respectively. 

\subsection{discussions}

\begin{figure}[htb]
\null\vspace{1cm}
\centerline{\epsfig{file=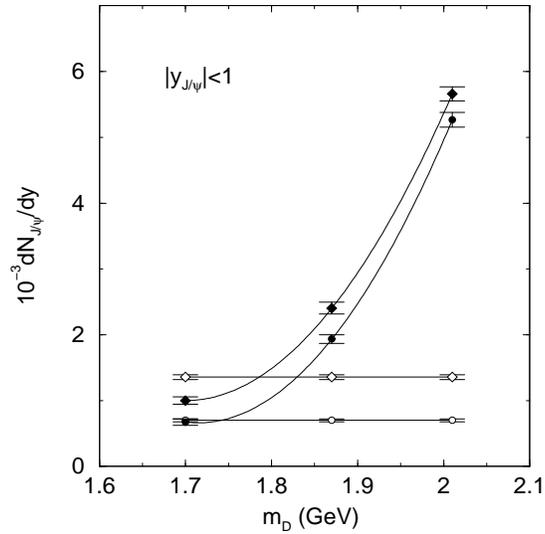,width=2.8in,height=2.8in,angle=-90}}
\null\vspace{0.5cm}
\caption{Number of produced $J/\psi$ per unit rapidity 
with $|y_{J/\psi}|<1$ as a function
of $D$ meson mass. Open symbols are for production in the
parton phase while filled symbols are for the final number including 
production from the hadron phase. Circles and diamonds are, respectively, 
the results with and without color screening in the parton phase.
Error bars are statistical.}
\label{fig9}
\end{figure}

The relative importance of $J/\psi$ production from the parton and 
hadron phases is more clearly seen in Fig.~\ref{fig9}, where we show the number 
of produced $J/\psi$ per unit rapidity with $|y_{J/\psi}|<1$ 
as a function of the $D$ 
meson mass. The results with color screening in the parton phase 
are shown by open circles for production from the parton phase and
filled circles for the final $J/\psi$ number including production
from the hadron phase. Without color screening in the parton phase, 
the corresponding results are given by diamonds. As expected, 
including color screening effect leads to a decrease 
in the final $J/\psi$ number. As the $D$ meson mass increases, 
the final $J/\psi$ yield increases strongly due to production from the
hadron phase, and the effect due to color screening becomes less
important. The results using $m_D=2.01$ GeV give the upper bound
for $J/\psi$ production while those using $m_D=1.87$ GeV give 
a lower bound. We note again that the net effect of a reduced 
$D$ meson mass of $m_D=1.70$ GeV is the destruction of 
$J/\psi$'s produced from the parton phase in the absence of
color screening. 

\begin{figure}[htb]
\null\vspace{1cm}
\centerline{\epsfig{file=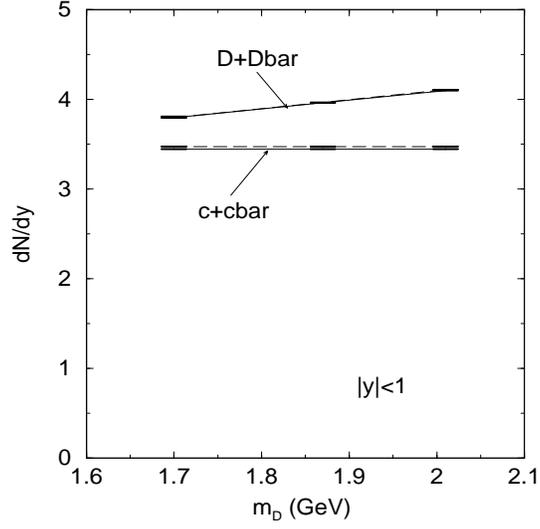,width=2.8in,height=2.8in,angle=-90}}
\null\vspace{0.5cm}
\caption{Midrapidity densities of charm quarks before (lower dashed line)
and after (lower solid line) parton evolution as well as of charm
mesons before (upper dashed line) and after (upper solid line) hadron
evolution.}  
\label{fig10}
\end{figure}

Part of increased production of $J/\psi$ in midrapidity when
the effective $D$ meson mass increases is due to increase in the
charm meson rapidity density during hadronization. This is shown
in Fig.~\ref{fig10}, where we give the midrapidity densities of charm quarks 
before (lower dashed line) and after (lower solid line) parton evolution 
as well as of charm mesons before (upper dashed line) and after 
(upper solid line) hadron evolution.
During parton evolution, the rapidity density of charm 
and anti-charm quarks decreases slightly due to collisions with
gluons. Hadronization increases the midrapidity density
of resulting charm mesons, and its effect increases with 
increasing charm meson effective mass. The midrapidity density 
of $D$ and $\bar{D}$ mesons is seen to increase by 8\% when $m_D$
increases from 1.70 GeV to 2.01 GeV. However, this would 
lead to at most 20\% increase in the $J/\psi$ midrapidity
density and cannot account for the more than factor of 5 increase
in $J/\psi$ production shown in Fig.~\ref{fig9}. The dominant effect for such
a larger increase is thus due to the reduced threshold for $J/\psi$ 
production when $D$ meson mass increases.

\begin{figure}[htb]
\null\vspace{1cm}
\centerline{\epsfig{file=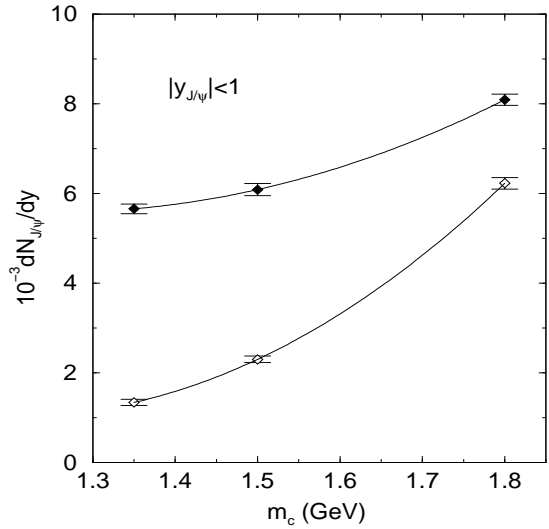,width=2.8in,height=2.8in,angle=-90}}
\null\vspace{0.5cm}
\caption{Midrapidity density of $J/\psi$ as a function of the charm 
quark mass for the production in the parton phase (open diamonds) and
the final number including the production in the hadron phase
(filled diamonds). Error bars are statistical.}
\label{fig11}
\end{figure}

The charm quark mass strongly influences the number of $J/\psi$ 
produced from the parton phase. Fig.~\ref{fig11} shows the $J/\psi$ rapidity 
density as a function of the charm quark mass. To get the upper
limit of $J/\psi$ production, the $D$ meson mass is set to be
2.01 GeV and color screening is not included. The $J/\psi$ number increases
by a factor of 72\% when the charm quark mass 
increases from the value 1.35 GeV used in PYTHIA to 1.5 GeV used in 
the constituent quark model.  If the charm quark mass is taken to be
1.8 GeV, we observe about a factor of 4 increase in $J/\psi$
production from the parton phase compared with the $m_c=1.35$ GeV case. 

In the kinetic model approach of Ref. \cite{rthews1}, the midrapidity 
density of charm quarks in Ref. \cite{rthews1} is
about 2.5 because the largest rapidity range for about 10 pairs of 
charm and anti-charm quarks is 4 in that study. This is larger than that
shown in Fig.~\ref{fig10}, which is about 3.48/2=1.74. Since $J/\psi$
production is roughly proportional to the square of the number of 
charm quarks, this can lead to about a factor of 2 difference
between our results and those from Ref. \cite{rthews1}. 
On the other hand, the cross section we use for $J/\psi$ production 
from the charm and anti-charm quarks is about a factor two larger 
than the peak value used in the kinetic model. The effects due
to different midrapidity charm quark densities and $J/\psi$
production cross sections in these two studies are thus largely
cancelled out. The different conclusions of $J/\psi$ suppression
in our study and $J/\psi$ enhancement in the kinetic approach
are results of different heavy ion collision dynamics used in
the two studies. In our approach, the parton phase is generated
from the mini-jet gluons from initial hard and semi-hard collisions 
between nucleons and is evolved using the transport model.
In the kinetic approach, an equilibrated quark-gluon plasma is 
assumed to be formed in the initial stage and later expands according
to the 1-D Bjorken hydrodynamical model \cite{jbjorken1}. 
Further studies are required
to determine the validity of these predictions.

\section{Summary}
\label{summary}

In this paper, we have studied the effects due to  
interactions between charm and anti-charm quarks in the initial 
parton phase and between charm mesons in the
final hadron phase on $J/\psi$ production in 
central Au+Au collisions at RHIC. Using a multi-phase transport 
model, we find that because of the high density of charm quarks 
at early times, the $J/\psi$ can be produced from the interactions of 
charm quarks. This is consistent with the conclusion of the
kinetic formation model. However, the more realistic space-time
evolution given by the transport model leads to an overall 
suppression of $J/\psi$ particles in the parton phase,
instead of an enhancement predicted by the kinetic model.

We have also shown that $J/\psi$ can be produced in the hadron
phase at RHIC. This is due to the large $D$ meson mass compared with
charm quark mass. Although we have not taken into account
parton-hadron interactions in the multi-phase model, both parton phase 
and hadron phase can exist in the same space-time region. 
This allows the production of $J/\psi$'s in the mixed phase. 
At RHIC energies, the multi-phase model shows that hadron
medium effect may lead to a net destruction of the $J/\psi$ produced 
in the parton phase. In addition, color screening is shown to have 
significant effect on the final $J/\psi$ yield.

In the present study, we have not included the effect due to  
gluon shadowing \cite{nhammon1}, which would reduce the charm quark 
and $J/\psi$
numbers. As the multi-phase transport model predicts an overall suppression
of the directly produced $J/\psi$, including gluon shadowing effect on
the initial charm quark distribution will not change qualitatively
the conclusion from the multi-phase model. Also, $J/\psi$ from 
decays of $\psi^\prime$ or $\chi_c$ was not considered in our study. 
The contribution from primary $\psi^\prime$ or $\chi_c$ to
final $J/\psi$ are negligible 
due to their large destruction cross sections compared 
with that for $J/\psi$. Although the production cross sections for
$\psi^\prime$ and $\chi_c$ from charm quark interactions and charm meson
interactions are expected to be larger than those for $J/\psi$, the
thresholds for these reactions are also higher. In particular,
the statistical fragmentation model 
\cite{pbraunmunzinger1,pbraunmunzinger2} indicates that 
the feeddown should be small for central collisions. We thus 
expect that our conclusion of $J/\psi$ suppression in heavy ion 
collisions at RHIC will not be qualitatively modified. \\

\acknowledgments

We thank P. Braun-Munzinger, M. Gyulassy, J. Kapusta, S. Pratt, 
R. Rapp, A.T. Sustich, 
C. Teal, R. Thews, W. Weise, and I. Zahed for helpful 
discussions. We also thank the Parallel Distributed System
Facility at the National Energy Research Scientific Computer
Center for providing computer resources. This work is supported
by the U.S. National Science Foundation under Grant No. 0088934 and
Grant No. 0098805, the Arkansas Science and Technology Authority
under Grant No. 01-B-20, the Welch Foundation under Grant No. A-1358,
and the Texas Advanced Research Program under Grant No. FY99-010366-0081.

\end{document}